\documentclass[11pt]{article}

\usepackage{graphicx}
\usepackage{amssymb}
\usepackage{amsmath}

\begin{document} 
\title{ Time and foreign exchange markets }
\author{Luca Berardi$^\star$ and Maurizio Serva$^\dagger$}
\date{}
\maketitle

\begin{small}
\begin{tabular}{cc}
${}^\star \,$ Dipartimento di Ingegneria Elettrica &
${}^\dagger \,$ Dipartimento di Matematica and I.N.F.M. \\
Universit\`a degli Studi, L'Aquila & 
Universit\`a degli Studi, L'Aquila \\
67040 Poggio di Roio (AQ) - Italy &
67010 Coppito (AQ) - Italy
\end{tabular}
\end{small}
\bigskip



\begin{abstract}

The definition of time is still an open question when one deals with
high frequency time series.  If time is simply the calendar time,
prices can be modeled as continuous random processes and values
resulting from transactions or given quotes are discrete samples of this
underlying dynamics.  On the contrary, if one takes the business time
point of view, price dynamics is a discrete random process, and time
is simply the ordering according which prices are quoted in the
market.  In this paper we suggest that the business time approach is
perhaps a better way of modeling price dynamics than calendar time.
This conclusion comes out from testing probability densities and conditional 
variances predicted by the two models against the experimental ones.
The data set we use contains the DEM/USD 
exchange quotes provided to us by Olsen \& Associates during a period of one
year from January to December 1998. In this period 1,620,843 quotes
entries in the EFX system were recorded. 

\end{abstract}



\bigskip
\bigskip
\bigskip
\bigskip
\bigskip
\bigskip
\bigskip
Corresponding author: 

Maurizio Serva, email: serva@univaq.it,
tel. +390862433153, fax. +390862433180

\newpage

\section{Introduction}
                                      
In the high-frequency arena there are two main-streams
about modeling the stochastic properties of quotes.
The first approach is to consider quotations as sampled values
of an underlying continuous-time random process \cite{Merton90},
\cite{Nelson90}. Sampling is itself a random operation, thus
introducing a twofold uncertainty in the price determination
\cite{Mainardi00}, \cite{Scalas00}. In this framework, time in the
model flows continuously, and is called {\bf calendar time}.

\noindent In the second approach, quoted prices are modeled through a
discrete-time stochastic process \cite{Taylor86}; in this setting,
time is just the natural total order relation among quotations, and it
is iso-morphic with the set of non-negative integers (being time $0$,
the time associated to the first considered quotation). This is the
{\bf business time} approach, and randomness only enters in the
determination of prices.  It should be pointed out, however, that the
waiting times between two quotes are also random quantities, but they
are assumed to not contribute to the price determination process.

Whether a calendar-time or a business-time framework should be adopted
in modeling the stochastic nature of financial quotes, has been a
longly debated issue by the finance research community, and it clearly
depends on many factors, like, for example: a) adherence to the
physical behavior of reported prices, b) usefulness in terms of a
theory to be developed, and c) last but not least, a matter of
taste. See, for example \cite{Baviera99}, \cite{Pasquini00},
\cite{Baviera01}.

In this paper, we suggest that business time is perhaps a better 
tool for modeling the asset dynamics than calendar-time.
In order to support our claim, we consider:    
1) returns corresponding to a given calendar time lag and any business time lag,  
2) returns corresponding to the same calendar time lag but having a fixed  
business time lag.
We find out that their statistical
properties are different consistently with the business hypothesis
and inconsistently with the calendar one.
In practice, we estimate some variances and some
probability densities
whose behavior is different in the two scenarios. 

The dataset we use contains the DEM/USD exchange quotes
taken from Reuters' EFX pages (the dataset having been supplied by
Olsen \& Associates) during a period of one year from January to
December 1998. In this period 1,620,843 quotes entries in the EFX
system were recorded.  The dataset provides a continuously updated
sequence of bid and ask exchange quotation pairs from individual
institutions whose names and locations are also recorded.
The reason for using FX data is that this market is not subject to
any working time restriction; in fact, it is open 24 hours a day, seven days a week.
This is in contrast to stock markets, where artificial time regulation
would have made more difficult, if not impossible, to find out
the results outlined in this paper.

\section{Business time vs Calendar Time}

\subsection{Calendar Time}

In the calendar time framework, prices are modeled as continuous-time
random processes. Clearly, market quotes are not defined for every $t
\in \mathbb{R}$, but only at discrete intervals, whose extensions in
time are called \emph{calendar lags} (usually ranging from 2sec. to
several minutes, sometimes hours).  Nevertheless, according to the
calendar time picture, prices are usually considered as discrete
samples of an underlying continuous-time random process.

The model of price dynamics in calendar time therefore has the
following structure:
\begin{equation}
S(t+\Delta) = S(t) e^{R_\Delta(t)}
\label{eq:CT-modelB}  
\end{equation}
where $S(t)$ and $S(t+\Delta)$ are the spot prices
at times $t$ and $t+\Delta$,
$\Delta$ is an arbitrary calendar time lag and
$R_\Delta(t)$ is the {\it aggregated} return of 
prices over the time interval $[t,t+\Delta]$.

Considering a framework where prices evolve over the calendar time, it
is generally assumed that quotes result from a random sampling at
times $t_0, \ldots t_n$ of the continuous-time underlying process
$S(t)$. In a pure calendar time framework such a random sampling is
uncorrelated with the process $S(t)$ itself. We observe however that
this is only valid as an approximation; indeed, several studies have
shown a weak correlation between the sequence of lags and that of
returns, among which we cite \cite{Raberto02}.

The last assumption usually made in order to complete the model
description in the calendar time setting is that the variance of
$R_\Delta(t)$ is a linear function of the 
calendar time lag $\Delta$, i.e.:
\begin{equation}
Var[R_\Delta(t)] = \sigma^2 \Delta
\label{eq:CT-variance}
\end{equation}
If the logarithm of $S(t)$ has independent increments the above 
equation obviously holds and $\sigma$ is the constant volatility.  
However, it is well known that independence does not hold 
because of volatility clustering which is due to the correlation 
of the absolute values of returns\cite{Pasquini99}.
As a consequence, in spite of a constant volatility, one has a time 
dependent volatility.
Nevertheless, the above behavior of the variance 
still holds true but $\sigma^2$ 
is now the average of the squared volatility.  
For our purposes we only assume that the above equality holds
and we do not need of specific assumptions concerning volatility behavior. 

Let us define the process $M(t)$ as follows:
\begin{equation}
M(t+\Delta) = M(t) + M_\Delta(t)
\end{equation}
where $M_\Delta(t)$ represents the number of given quotes (samples) in the
interval $[t,t+\Delta]$. Clearly, $M(t)$ is a non-decreasing random
process assuming integer values. We also observe that $M(t)$ as a
function of $t$ is piecewise constant, and its value increases by one
each time a quote is given (i.e. at times $t_0, \ldots t_n$).

Given the assumptions made so far, it follows that the process $M$ and
the process $S$ are mutually independent. Hence, it follows that
the probability density of returns corresponding 
to a calendar time lag $\Delta$ is insensitive from the condition 
that $M_\Delta(t)$ is also fixed to a value $m$.
In symbols:  
\begin{equation}
P[R_\Delta(t) | M_\Delta(t) = m] = P[R_\Delta(t)]
\label{eq:CT-cond_density}
\end{equation}
and, in particular, the associated variance exhibits the same
insensitiveness:
\begin{equation}
Var[R_\Delta(t) | M_\Delta(t) = m] = Var[R_\Delta(t)] =
\sigma^2 \Delta
\label{eq:CT-cond_variance}
\end{equation}

\noindent Therefore, we can summarize the calendar time
hypothesis as follows:

\paragraph{Hypothesis H1:}
The asset prices evolve over calendar time, i.e. according to the
model in  Eq.~(\ref{eq:CT-modelB}) and Eq.~(\ref{eq:CT-variance}) holds. 
Moreover the processes $S$ and $M$ are mutually independent, therefore
Eq.~(\ref{eq:CT-cond_density}) and Eq.~(\ref{eq:CT-cond_variance}) hold.

\bigskip

Let us anticipate that the main argument of the paper
is based on the estimation of the quantities in
Eq.~(\ref{eq:CT-cond_density}) and Eq.~(\ref{eq:CT-cond_variance}).  
We will show with enough evidence that the two equalities are 
largely violated in a way which, on the contrary, is consistent
with the business framework. 

\subsection{Business Time}

In the business-time approach, price dynamics is modeled as a
discrete-time random process. Indeed, the time basis is the ordered
sequence of times at which prices are quoted in the markets. It is
therefore a set isomorphic with the set of non-negative integers.  In
such a framework the statistic model of price dynamics in the
business-time framework is the following:
\begin{equation}
S(n+m) = e^{R_m(n)}S(n)
\label{eq:BT-model}
\end{equation}
where $S(n)$ and $S(n+m)$ are 
the asset price at business times $n$ and $n+m$ while 
$R_m(n)$ is the \emph{aggregated} return over $m$ consecutive quotes.
It is then clear that the only time-dependence affecting the price process is
based on the global ordering of events while the
return is independent from calendar lag.
Notice that we refer to $m$ as
the business time lag as opposed to the calendar time lag
$\Delta$ introduced in the previous section.

Considering the price dynamics in a business time setting naturally
leads to the following assumption:
\begin{equation}
Var[R_m(n)] = \hat{\sigma}^2 m
\label{eq:BT-variance}
\end{equation}
whose motivation is the same of that provided for the analogous
assumption in the calendar time hypothesis.
We also define the random process:
$$
T(n+m) = T(n) + T_m(n) 
$$ where $T(n)$ is the stochastic calendar time at 
business time $n$ and $T_m(n)$ corresponds
to the calendar lag $T(n+m) - T(n)$, i.e. the time
elapsed from $T(n)$ after the occurrence of $m$ consecutive quotes.  
It can be readily seen that there is a direct
connection between $T(n)$ and the process $M(t)$ defined in the
previous subsection. In fact, $M(t)=n$ with $t \in [T(n),T(n+1))$,
and, moreover, the following relation holds:
$$
M_{T_m(n)}(T(n)) = m
$$ 
for an arbitrary positive integer $m$.

\bigskip
Given the assumption of statistical independence between $S(n)$ and
$T(n)$, for a generic $\Delta$ the following relation holds:
\begin{equation}
P[R_m(n) | T_m(n) \in [\Delta-\epsilon, \Delta+\epsilon] ] = P[R_m(n)] 
\label{eq:BT-pdf}
\end{equation}
where $\epsilon$ is a fixed quantity.
The above equation  states that the probability density of returns corresponding to a 
business time lag $m$ is insensitive to the condition that
the calendar time lag is also fixed to a value around $\Delta$.
In particular we have for the variance:
\begin{equation}
Var[R_m(n) | T_m(n) \in [\Delta-\epsilon, \Delta+\epsilon] ] = Var[R_m(n)]
 = \hat{\sigma}^2 m
\label{eq:BT-cond_variance}
\end{equation}
which is the business time analogue of
Eq.~(\ref{eq:CT-cond_variance}).

Given all the assumptions made so far, we are ready to formulate the
hypothesis of prices dynamics in a business time setting.

\bigskip
\paragraph{Hypothesis H2:}
Asset prices follow the model in Eq.~(\ref{eq:BT-model}) and
Eq.~(\ref{eq:BT-variance}) holds. Moreover, the processes $S$ and $T$
are independent, it follows that Eq.~(\ref{eq:BT-pdf}) and
Eq.~(\ref{eq:BT-cond_variance}) hold.

\bigskip

Before concluding this preliminary outline of the two basic approaches
used to describe price dynamics (i.e. calendar time \& business time)
we also give another important property of some of the quantities
involved so far, which will turn useful in the remaining part of the
paper. \\ With all the positions previously made, let us first observe
that the following relation holds:~\footnote{This follows from the
stationarity of the process $M_\Delta(T(n))$. In particular:
$E[M_\Delta(T(n))]$ does not depend on $T(n)$ so we drop the
sub case. Moreover, $E[M_{k \Delta}] = k E[M_\Delta]$, since the
average number of quotes in $k$ intervals of the same length sums up
to $k$ times the value for the single interval, from which the
proportionality follows.}
$$
E[M_\Delta(T(n))] = \alpha \Delta
$$
for a suitable constant $\alpha$. Simply put, this property states
that the expected value of the number of quotes in an interval
$\Delta$ is proportional to $\Delta$ itself.

Finally, considering the composition of the price process in business
time and the process representing the number of quotes in a
given calendar time lag $\Delta$, it can be shown that:
\begin{equation}
Var[R_{M_\Delta(t_n)}(n)] = \hat{\sigma}^2 E[M_\Delta(T(n))] =
\hat{\sigma}^2 \alpha \Delta
\label{eq:BT-CT-crossover}
\end{equation}
Thus, in the business time hypothesis, we also expect the variance in
(\ref{eq:BT-CT-crossover}) to be proportional to $\Delta$.

As already anticipated, all equalities in this subsection
are supported by the following statistical analysis  confirming
the validity the business time framework.

\section{Statistical Estimators}

In this and next section we carry out some experimental
analysis in order to best fit the description of prices dynamics
choosing between the two distinct possibilities concisely modeled
by hypotheses H1 and H2.

In this section, in particular, we will define some statistical
estimators, i.e. functions of the data contained in high frequency
time series, and relate them to their probabilistic counterparts
defined in the previous section.

Our dataset refers to the FX ratio USD/DM over the whole year 1998
and the price $S_i$ we consider in this
paper is the half sum of bid and ask (mid-price) while $t_i$ denote
the time at which the $i$-th price is given.  Some automatic filtering
procedure is also applied, to remove erroneous recording, which we are
able to individuate since they correspond to prices macroscopically
different from previous and subsequent ones.

\bigskip
Let $\mathcal{R} = \left\{ r_i \right\}_{i=0,1,\ldots,L}$ be the
series of elementary returns $r_i$ defined as:
$$
r_i = \log \frac{S_{i+1}}{S_i} \qquad i=0,1,\ldots,L
$$
and let  $\mathcal{T} =
\left\{ \tau_i \right\}_{i=0,1,\ldots,L}$
be the series of temporal lags defined as: $\tau_i=t_{i+1}-t_i$.

Now consider the series $\mathcal{R}(\Delta,m) = \left\{ r_i(\Delta,m)
\right\}_{i=0,1,\ldots,L(\Delta,m)}$; the $r_i(\Delta,m)$ are obtained
by summing $m$ consecutive elementary returns (where $m$ is fixed) and
subsequently retaining only the  $L(\Delta,m)$
sums corresponding to a lag in the
interval $[\Delta-\epsilon, \Delta+\epsilon]$ (i.e. the sum of the
corresponding $m$ elementary lags $\tau_i$ is in the interval
$[\Delta-\epsilon,\Delta+\epsilon]$, where $\epsilon$ is also a fixed
quantity).

The mean and variance of such a series are respectively defined as:
\begin{eqnarray*}
\mu(\Delta,m) &=& \frac{1}{L(\Delta,m)} \sum_{i=1}^{L(\Delta,m)}
r_i(\Delta,m) \\
v(\Delta,m) &=& \frac{1}{L(\Delta,m)} \sum_{i=1}^{L(\Delta,m)}
[r_i(\Delta,m) - \mu(\Delta,m)]^2
\end{eqnarray*}
We observe that $v(\Delta,m)$ represents an estimation of the quantity
$Var[R_\Delta(t) | M_\Delta(t)=m]$ for the calendar time model; and,
as pointed out before, we expect it to be a linear function of
$\Delta$, should hypothesis H1 be correct.
Moreover, in this hypothesis, we expect this variance to be constant 
with respect to $m$ if $\Delta$ is fixed.

Alternatively, considering the business time framework, $v(\Delta,m)$
can also be seen an estimator of the quantity $Var[R_m(n) | T_m(n) \in
[\Delta-\epsilon, \Delta+\epsilon] ]$ defined in
Eq.~(\ref{eq:BT-variance}); should hypothesis H2 be correct we expect, 
given $m$, that $v(\Delta,m)$ is approximately constant with respect to 
$\Delta$.
Moreover, in this hypothesis, we expect this variance to be linear 
in $m$ even if $\Delta$ is fixed.

With the same set of data  $\mathcal{R}(\Delta,m)$ 
we can can compute the empirical pdf
of returns with fixed $\Delta$ and with fixed $m$.
This pdf is an estimator of $ P[R_\Delta(t) | M_\Delta(t) = m] $
and also of 
$P[R_m(n) | T_m(n) \in [\Delta-\epsilon, \Delta+\epsilon] ]$.

\bigskip
Consider now the series $\mathcal{R}(\Delta) = \left\{ r_i(\Delta)
\right\}_{i=0,1,\ldots,L(\Delta)}$ obtained from $\mathcal{R}$ by
summing consecutive elementary returns until the corresponding lag
becomes equal or greater than $\Delta$.
The number of the elements of this series is $L(\Delta)$ and
the mean and variance are respectively defined as:
\begin{eqnarray*}
\mu(\Delta) &=& \frac{1}{L(\Delta)}
\sum_{i=1}^{L(\Delta)} r_i(\Delta) \\
v(\Delta) &=& \frac{1}{L(\Delta)}
\sum_{i=1}^{L(\Delta)} [r_i(\Delta) -
\mu(\Delta)]^2
\end{eqnarray*}
In the calendar time framework, $v(\Delta)$ estimates the quantity
$Var[R_\Delta(t)]$, defined in Eq.~(\ref{eq:CT-variance}).  In the
business time case, instead, $v(\Delta)$ estimates the quantity
$Var[R_{M_\Delta(t_n)}(n)]$ in Eq.~(\ref{eq:BT-CT-crossover}). In both
cases we expect this quantity to grow linearly with $\Delta$.

With the same set of data  $\mathcal{R}(\Delta)$ 
we can can compute the empirical pdf
of returns with fixed $\Delta$ (any $m$).
This pdf is an estimator of $ P[R_\Delta(t)] $.


\section{The choice of the correct model from data analysis}
\label{sec:test_results}

\begin{figure}
\begin{center} 
\includegraphics[width=0.7\textwidth,angle=-90]{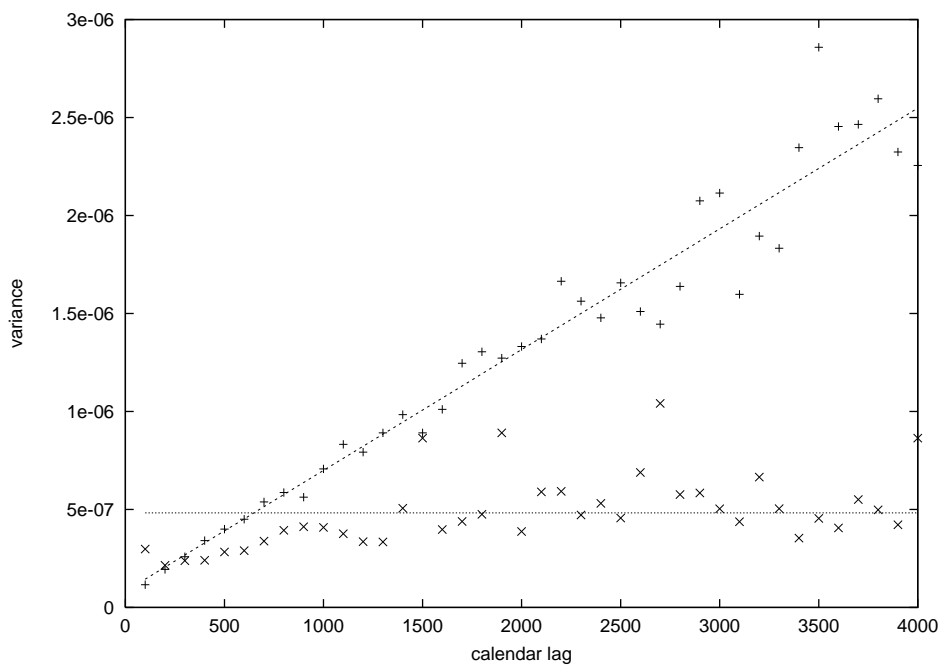}
\caption{ \footnotesize We plot here the statistical estimators
$v(\Delta)$ (+ symbols) and $v(\Delta,m)$ with $m=40$  ($\times$ symbols)
for different values of the calendar time lag $\Delta$. It can be
readily seen that while $v(\Delta)$ varies linearly with $\Delta$, the
quantity $v(\Delta,m)$ is approximately constant. 
Therefore, if the business time lag is fixed (at $m=40$), 
the variance of the returns does not scale with time lag $\Delta$. 
This would indicate that business time lag rather than
calendar time lag forms the important independent variable. 
A linear fit was
computed in the first case resulting in $v = 6.16E-10 \Delta + 8.26E-8
\quad$ and a constant fit in the second resulting in $v = 4.83E-7$.}
\label{fig:variance_Delta}
\end{center}
\end{figure}

We have now sufficient information in order to accept or discard
hypothesis H1 and H2, as a result of an empirical data analysis.

First, we have computed the statistical estimators $v(\Delta)$
and $v(\Delta,m=40)$ 
as defined in the previous section and both plotted
in Fig.~\ref{fig:variance_Delta} for different values of the
calendar time lag $\Delta$. It can be readily seen that while
$v(\Delta)$ varies linearly with $\Delta$, the quantity
$v(\Delta,m)$ is approximately constant. 
Indeed, a linear fit was computed in the first case
resulting in $v = 6.16E-10 \Delta + 8.26E-8 $ 
and a constant fit in the second resulting in
$v = 4.83E-7$.

\begin{figure}
\begin{center} 
\includegraphics[width=0.7\textwidth,angle=0]{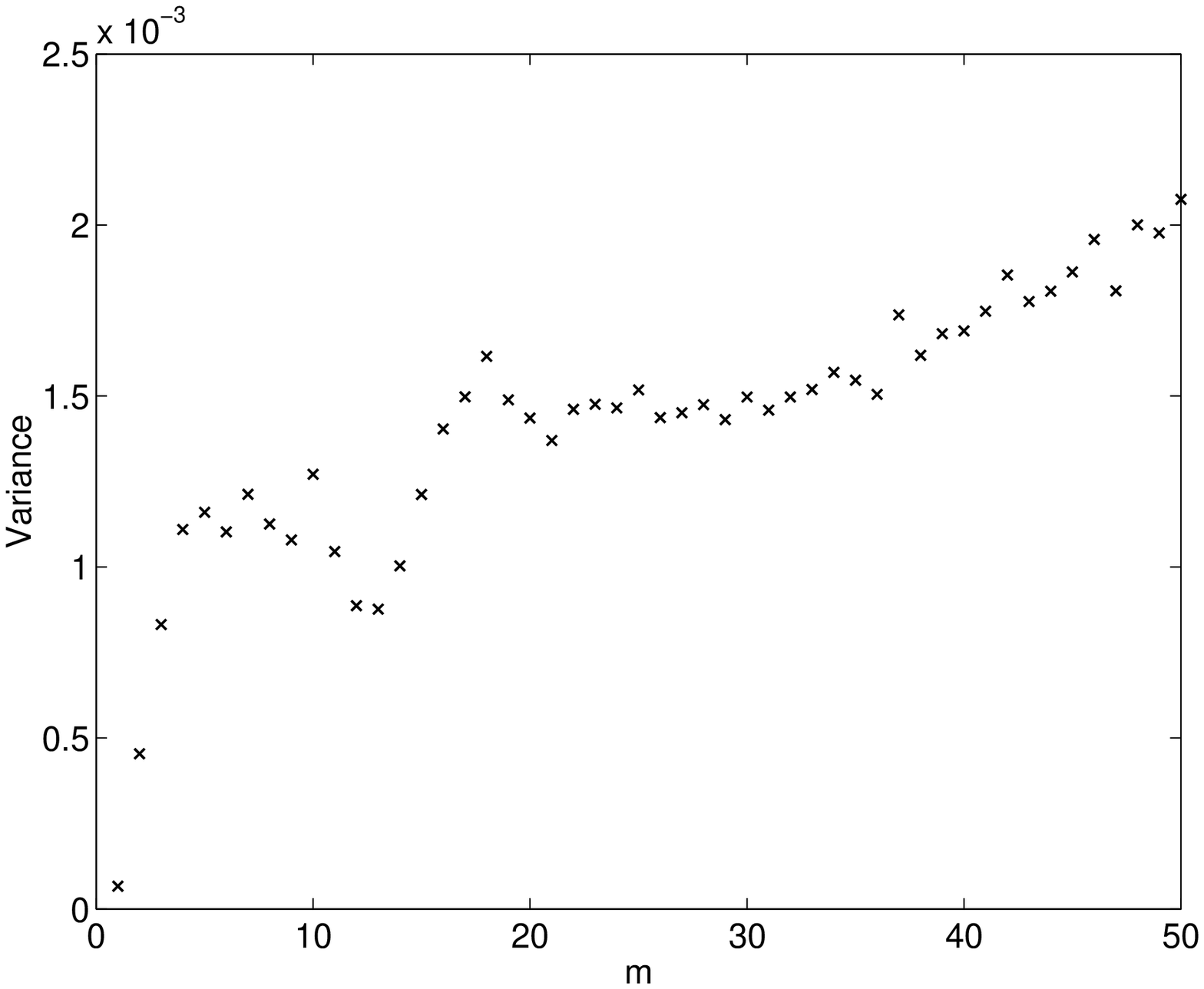}
\caption{ \footnotesize We plot here the statistical estimator
$v(\Delta,m)$ with a fixed $\Delta=1000 \pm 50$ 
for different values of the business time lag $m$. 
I can be seen that  $v(\Delta,m)$ grows with $m$
(even if not linearly in all range considered).}
\label{fig:variance_m}
\end{center}
\end{figure}

We recall that, according to the calendar time hypothesis the two
lines should be equal and proportional to $\Delta$, while in the
business time case the former should be proportional to $\Delta$,
while the latter should be constant. The
corresponding graphs in fig.~\ref{fig:variance_Delta} 
seem to suggest that the business time model 
is more likely valid, while the hypothesis of 
calendar time dynamics seems to be unlikely.
The same kind of behavior can be found if one
chooses the value of $m$ in a range between 5 and 100.

In fig.~\ref{fig:variance_m} we also plot the statistical 
estimator $v(\Delta,m)$  versus $m$
with a fixed $\Delta=1000 \pm 50$.
According to the calendar time hypothesis 
this quantity  should be constant
while, according to the business time hypothesis ,
should grow linearly in $m$. 
The behavior is not linear in all range
but, anyway, $v(\Delta,m)$ grows 
with respect to $m$, which also supports
the business time hypothesis.
It should be noticed that the choice of other values 
of the fixed $\Delta$ would not alter this picture.

\begin{figure}
\begin{flushleft} 
\includegraphics[width=0.7\textwidth,angle=-90]{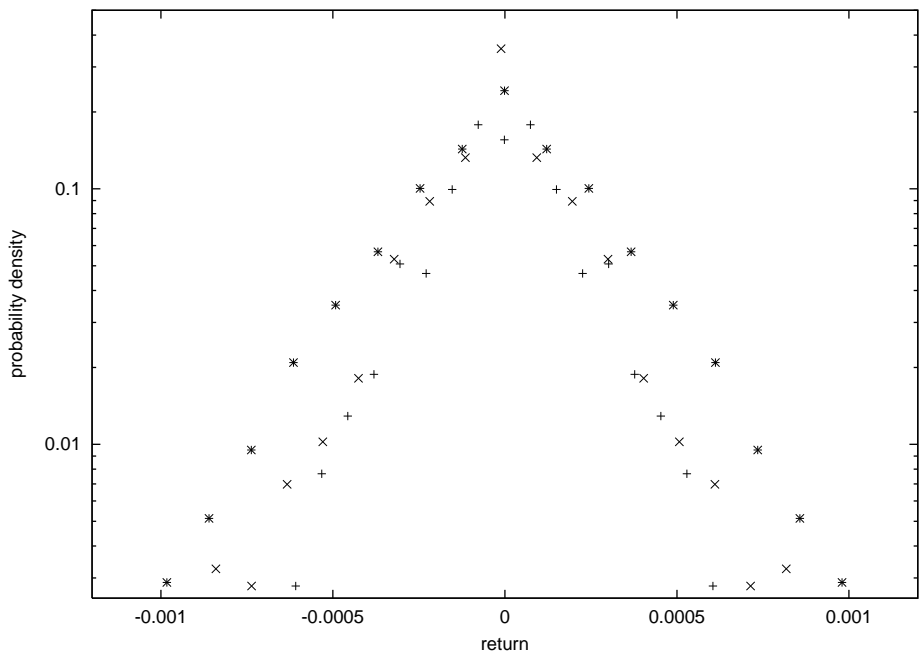}
\caption{ \footnotesize Estimated probability density functions for
$\mathcal{R}(\Delta=2sec) = \mathcal{R}(\Delta=2sec,m=1)$,
$\mathcal{R}(\Delta=100sec)$ and $\mathcal{R}(\Delta=100sec,m=1)$ in a
log-linear plot.  The first two pdf (+ symbols) coincide because of the
data set characteristics as explained in the text; the pdf for
$\mathcal{R}(\Delta=100sec,m=1)$ ($\times$ symbols) is roughly the same
of the first two while the pdf for $\mathcal{R}(\Delta=100sec)$
(star symbols) is macroscopically different having larger moments. 
The significance of the plots lies in the fact that
if $m=1$, then a large calendar time seems to make no difference,
whereas if $m$ is allowed to vary, then the PDF becomes fat, 
due to return aggregation. }
\label{fig:densities}
\end{flushleft}
\end{figure}

\bigskip
Second, we consider two distinct series of returns 
$\mathcal{R}(\Delta,m)$ and $\mathcal{R}(\Delta)$ 
(respectively $a$ and $b$) as defined in the previous section.

Since the minimum lag between two consecutive quotes is equal to $2$
seconds in the given database, the two series $a$ and $b$ coincide
for $\Delta=2$ sec; formally: $\mathcal{R}(\Delta=2 sec, m=1) =
\mathcal{R}(\Delta=2 sec)$.

We have subsequently compared the estimated probability density
functions (pdf) for the series $a$ and $b$ and the results are shown
in fig.~\ref{fig:densities}.

The figure is a log-linear plot of different probability densities,
For $\Delta=2 sec$ the pdf the two cases $\mathcal{R}(\Delta=2sec)$
and $\mathcal{R}(\Delta=2sec,m=1)$ 
exactly coincide because of the data set
characteristics as just explained.  For $\Delta=100 sec$ we observe a
remarkable difference between the pdf for the series
$\mathcal{R}(\Delta=100sec,m=1)$ and $\mathcal{R}(\Delta=100sec)$.
The former, in fact, is roughly the same as
$\mathcal{R}(\Delta=2sec)$, while the second is fatter (larger
moments). 

This fact disagrees with Eq.~(\ref{eq:CT-cond_density})
which is a consequence of calendar time hypothesis.
In fact, according to this equation the two pdf 
corresponding to $\mathcal{R}(\Delta=100sec,m=1)$ 
and $\mathcal{R}(\Delta=100sec)$ should be equal. 

On the contrary, one can immediately see that this result is 
in accordance with Eq.~(\ref{eq:BT-pdf}) and, therefore, 
with business time hypothesis.
In fact, $\mathcal{R}(\Delta=100sec,m=1)$ and 
$\mathcal{R}(\Delta=2sec,m=1)$ are roughly the same.
This experimental equality simply means that given the value of $m$
returns are substantially insensitive to $\Delta$
as stated in Eq.~(\ref{eq:BT-pdf}). 

In conclusion, this experimental result
provides further evidence that the correct model should be the one of
the process evolving over business time (hypothesis H2).

\section{Conclusions}

In this paper we suggest that the business time approach is perhaps a
better way of modeling price dynamics than calendar time.
In order to derive some insight from data we neglect
possible autocorrelation between returns and possible autocorrelation
between lags assuming implicitly that they would only give a second
order correction to our findings.  With this simplification our
results altogether seem to provide enough evidence for the rejection
of hypothesis H1 (calendar time model) and the acceptance of
hypothesis H2 (business time model). 
Nevertheless, it should be noticed that hypothesis H1
assumes that the sampling process is independent of the price evolution.
Therefore, our results do not rule out the continuous time model, but rather
they show that the the continuous time model would require
correlations between processes $M$ and $S$ to in order to fit the data.

The deep reason of the behavior we point out in this paper is that
when an asset (at least a forex asset) is not traded, 
the prices evolution is
slow while the evolution is fast when the asset is heavily traded.
A faster evolution corresponds to a larger volatility in calendar time
\cite{Dacorogna01,Lillo02}, therefore, one could even maintain 
the calendar point of view, but in this case
it should accept a seasonal modulation of volatility.  
The fact that the evolution of a price is slow
when there are few transactions
is very well known to practitioners but it is
still not accepted in its extremal consequence that prices are frozen
when assets are not traded at all.  
This is because this behavior is
in contrast to the stock market experience where opening prices are
different from previous night closing prices.  Nevertheless the
difference between the two markets is not astonishing if one thinks
that the stock market is artificially time regulated, while the forex
exchange market is an over the counter (OTC) market not subject to any
time restriction.

\bigskip

\bigskip

{\bf Acknowledgements}

We would like to thank Filippo Petroni for a number
of discussion on the subject and for his advice
about the manipulation of the high frequency data sets.



\bigskip


\begin{thebibliography}{99}


\bibitem{Baviera99} 
R. Baviera, M. Pasquini, M. Serva, D. Vergni and A. Vulpiani, 
{\it Correlations and multyaffinity in high frequency
financial data sets}, 
Physica A {\bf 300}, 551-557 (2001).

\bibitem{Baviera01} 
R. Baviera, M. Pasquini, M. Serva, D. Vergni and A. Vulpiani,
{\it Forecast in foreign exchange markets}, 
Eur. Phys. J. B {\bf 20} 473-479 (2001).

\bibitem{Dacorogna01} 
M. Dacorogna, R. Gensay, U. Maller, R. Olsen and O. Pictet, 
{\it An Introduction to High-Frequency Finance},
Academic Press. 2001. 

\bibitem{Lillo02} 
F. Lillo, J.D. Farmer and R. Mantegna, 
{\it Muster curve for price-impact function},
Nature {\bf 421}, 129-130 (2003)

\bibitem{Mainardi00} 
F. Mainardi, M. Raberto, R. Gorenflo and E. Scalas, 
{\it Fractional calculus and continuous-time finance II: the
waiting-time distribution},
Physica A, {\bf 287}, 468-481 (2000).

\bibitem{Merton90} 
R.C. Merton, {\it Continuous-Time Finance},
Blackwell Publishers, 1992.

\bibitem{Nelson90} 
D. Nelson,
{\it  ARCH Models as Diffusion Approximations}, 
Journal of Econometrics, {\bf 45}, 7-38 (1990).

\bibitem{Pasquini99} 
M. Pasquini and M. Serva, 
{\it Multiscaling and clustering of volatility},  
Physica A, {\bf 269}, 140-147 (1999).

\bibitem{Pasquini00} 
M. Pasquini and  M. Serva,
{\it Indeterminacy in foreign exchange markets}, 
Physica A {\bf 277}, 228-238 (2000).

\bibitem{Raberto02} 
M. Raberto, E. Scalas, and F. Mainardi,
{\it Waiting-times and returns in high-frequency financial data: an
empirical study},
Physica A, {\bf 314}, 751--757 (2002).

\bibitem{Scalas00} 
E. Scalas, R. Gorenflo and F. Mainardi,
{\it  Fractional calculus and continuous-time finance}, 
Physica A {\bf 284}, 376-384 (2000).

\bibitem{Taylor86} 
S. Taylor,
{\it Modeling financial time series}, 
John Wiley \& Sons, New York, (1986).


\end{thebibliography}

\end{document}